\documentclass[11pt]{article}
\usepackage{amssymb}

\title{Kochen-Specker Theorem, Physical Invariance and Quantum Individuality}

\author{{\sc C. de Ronde}$^{1,2}$ and {\sc C. Massri}$^{3}$}
\date{}

\usepackage[margin=3.5cm]{geometry}

\begin{document}

\bibliographystyle{plain}
\maketitle

\begin{center}
\begin{small}
1. Philosophy Institute Dr. A. Korn (UBA-CONICET)\\
2. Center Leo Apostel for Interdisciplinary Studies\\ Foundations of the Exact Sciences (Vrije Universiteit Brussel).\\
3. Department of Mathematics (UBA-CONICET).\\
\end{small}
\end{center}

\begin{abstract}
\noindent In this paper we attempt to discuss what has Kochen-Specker (KS) theorem to say about physical invariance and quantum individuality. In particular, we will discuss the impossibility of making reference to objective physical properties within the orthodox formalism of quantum mechanics. Through an analysis of the meaning of physical invariance and quantum contextuality we will derive a Corollary to KS theorem that proves that a vector in Hilbert space cannot be interpreted coherently as an object possessing (objective) physical properties. As a consequence, the notion of quantum object can be only defined in terms of nomological properties. We conclude the paper by analyzing the consequences of this Corollary to KS theorem for the ongoing debate about quantum individuality.
\end{abstract}
\begin{small}


\end{small}

\newtheorem{theo}{Theorem}[section]

\newtheorem{definition}[theo]{Definition}

\newtheorem{lem}[theo]{Lemma}

\newtheorem{met}[theo]{Method}

\newtheorem{prop}[theo]{Proposition}

\newtheorem{coro}[theo]{Corollary}

\newtheorem{exam}[theo]{Example}

\newtheorem{rema}[theo]{Remark}{\hspace*{4mm}}

\newtheorem{example}[theo]{Example}

\newcommand{\proof}{\noindent {\em Proof:\/}{\hspace*{4mm}}}

\newcommand{\qed}{\hfill$\Box$}

\newcommand{\ninv}{\mathord{\sim}} 

\newtheorem{postulate}[theo]{Postulate}

\section{Introduction}

The interpretation of states (vectors in Hilbert space) in quantum mechanics (QM) remains, still today, one of the most controversial subjects of debate in both physics and philosophy of physics. As remarked in a recent paper \cite[p. 475]{PuseyBarrettRudolph12}: ``Quantum states are the key mathematical objects in quantum theory. It is therefore surprising that physicists have been unable to agree on what a quantum state truly represents.'' In this paper we will argue that the kernel of this debate stands, quite paradoxically, on a fundamental idea ---introduced by the first postulate of orthodox quantum theory--- which is simply untenable, namely, that a vector can be interpreted as the state of a physical (quantum) system or object. Unfortunately, this assumption has been taken for granted in many discussions and analysis regarding the interpretation of QM. In particular, it is assumed by Pusey, Barrett and Rudolph in order to prove their no-go theorem about the `reality of the quantum state'. According to the so called PBR theorem:

\begin{quotation}
\noindent {\small``[...] if the quantum state merely represents information about the real physical state of a system, then experimental predictions are obtained which contradict those of quantum theory. The argument depends on few assumptions. One is that a system has a `real physical state' not necessarily completely described by quantum theory, but objective and independent of the observer. This assumption only needs to hold for systems that are isolated, and not entangled with other systems. Nonetheless, this assumption, or some part of it, would be denied by instrumentalist approaches to quantum theory, wherein the quantum state is merely a calculational tool for making predictions concerning macroscopic measurement outcomes.'' \cite[p. 475]{PuseyBarrettRudolph12}}
\end{quotation}

\noindent We will show that this assumption should not only be denied by instrumentalist approaches but also by anyone who accepts or takes as a standpoint the orthodox formalism of QM. Jonathan Barrett has made the point that: ``People have become emotionally attached to positions that they defend with vague arguments, it's better to have a theorem.'' \cite[p. 157]{Reich12}  We certainly agree on both points. There is indeed a widespread misuse of basic physical concepts ---e.g. `state', `system', `physical reality', etc.--- responsible for a proliferation of undefined and vague problems within the foundational literature regarding QM. In order to expose this state of affairs we will put forward a physical argumentation based on the notion of invariance which will allow us to develop a No Go theorem for the existence of objective properties in QM. This theorem implies as a consequence that interpreting a vector in Hilbert space as `the physical state of a (quantum) system' is inconsistent with the mathematical formalism of QM. We will argue that this misinterpretation has played a significant role in stopping the theory from going beyond the metaphysics of actual entities and finding a coherent physical interpretation which explains what QM is really talking about.

The stance we will assume in order to develop our argument is that, while physics attempts to describe the world and physical reality, pure mathematics is a non-representational discipline which is not constrained by the physical world nor experience. Mathematics is a language which needs no metaphysical nor physical reference whatsoever. We take mathematics to be a discipline which remains at distance from both metaphysical and physical considerations, the only judge it accepts is its own internal coherency. Only when interpreted and related to experience, pure mathematics escapes the safety of rational speculation and deduction and becomes part of a physical theory. But in general, due to its non-representational character formal mathematical structures have no ``minimal interpretation'', or in other words, there is no ``self evident'' path to follow from a specific mathematical formalism into a physical theory (for a detailed discussion see \cite{daCostadeRonde16}).

The paper is organized as follows. In section 2, we analyze the orthodox formalism of QM and its metaphysical underpinnings which, we claim, appear from the choice of the concepts used in order to express the main postulates of the theory. Section 3 compares the invariance of classical theories to the contextual character of QM. In section 4 we continue to analyze the orthodox mathematical structure of QM and derive a No Go Theorem which proves that the interpretation of a vector in Hilbert space in terms of the state of a physical system which possess objective physical properties is inconsistent with the formalism of the theory. In section 5, we discuss the consequences of the No Go theorem with respect to several ongoing debates in the literature. Finally, in section 6 we present some final remarks.

\section{The Orthodox Quantum Formalism and its \\Metaphysical Underpinnings}

The orthodox formalism of QM is constructed within a vectorial Hilbert space. As we shall see in this section, this mathematical structure has been interpreted, from the very origin of the theory, following ---implicitly--- certain basic preconceptions of Newtonian physics. Although the formalism has shown many problems when interpreted under these lines, certain main ideas have remained present in all interpretations of QM. Hidden within language, the metaphysics of entities and actuality has prevailed up to the present. In the following section we shall firstly describe the mathematical structure of the theory, and secondly, we will expose the metaphysical underpinnings that have restricted the development of QM.

Let us begin by providing a mathematical account of the theory (for more on this see \cite{Conway}). First, we give the definition of Hilbert spaces and dimension using sets and functions. A \emph{complex vector space} ${\cal H}$ is a set with two algebraic structures satisfying the following axioms,

\begin{enumerate}
\item $x+(y+z)=(x+y)+z,\quad\forall x,y,z\in\mathcal{H}$
\item $x+y=y+x,\quad\forall x,y\in\mathcal{H}$
\item There exists $0\in\mathcal{H}$ such that $0+x=x,\quad\forall x\in\mathcal{H}$
\item There exists $-x\in\mathcal{H}$ such that $x+(-x)=0,\quad\forall x\in\mathcal{H}$.
\item $\lambda(x+y)=(\lambda x)+(\lambda y),\quad\forall \lambda\in\mathbb{C}, x,y\in  {\cal H}$.
\item $(\lambda_1+\lambda_2) x= (\lambda_1x)+(\lambda_2 x),\quad \forall \lambda_1,\lambda_2\in\mathbb{C}, x\in  {\cal H}$.
\item $(\lambda_1\lambda_2) x= \lambda_1(\lambda_2 x),\quad \forall \lambda_1,\lambda_2\in\mathbb{C}, x\in {\cal H}$.
\item $1x=x,\quad\forall x\in {\cal H}$.
\end{enumerate}

\noindent The most common example of a complex vector space is $\mathbb{C}^n$, where $n$ is a positive integer, $n\in\mathbb{N}$.
The elements $x\in {\cal H}$ of a complex vector space ${\cal H}$ are called \emph{vectors}, usually denoted $|x\rangle$.
A \emph{linear morphism} from a complex vector space ${\cal H}_1$ to another complex vector space ${\cal H}_2$ is
a function $f:{\cal H}_1\rightarrow {\cal H}_2$ such that

\begin{enumerate}
\item $f(x+y)=f(x)+f(y),\quad\forall x,y\in {\cal H}_1$.
\item $f(\lambda x)= \lambda f(x),\quad \forall \lambda\in\mathbb{C}, x\in {\cal H}_1$.
\end{enumerate}

\noindent  We say that the linear morphism $f$ is an \emph{isomorphism} if $f$ is bijective.
A \emph{finite-dimensional} complex vector space ${\cal H}$
is a complex vector space isomorphic to $\mathbb{C}^n$ for some $n\in\mathbb{N}$.
The number $n$ is called the \emph{dimension} of ${\cal H}$.
We say that the pair $({\cal H},\langle -|-\rangle)$ is a finite dimensional \emph{Hilbert space} if
${\cal H}$ is a finite dimensional complex vector space and $\langle -|-\rangle$ is an inner product in ${\cal H}$. An \emph{inner product} in ${\cal H}$ is a function $\langle -|-\rangle: {\cal H} \times {\cal H} \rightarrow \mathbb{C}$ satisfying,

\begin{enumerate}
\item $\langle x|y\rangle=\overline{\langle y|x\rangle},\quad\forall x,y\in {\cal H}$.
\item $\langle \lambda_1 x_1+\lambda_2 x_2|y\rangle = \lambda_1 \langle x_1|y\rangle +\lambda_2 \langle x_2|y\rangle ,\quad \forall \lambda_1,\lambda_2\in\mathbb{C}, x_1,x_2,y\in {\cal H}$.
\item $\langle x|x\rangle >0,\quad\forall x\neq 0, x\in {\cal H}$.
\end{enumerate}

\noindent  As an example, $\mathbb{C}^n$ with the usual inner product, $\langle v|w\rangle=\sum_{i=1}^n v_i\overline{w_i}$, is a Hilbert space of dimension $n$.

As we argued above, a mathematical formalism does not provide in itself an interpretation of its terms. In particular, a vector in Hilbert space does not have any self evident physical interpretation whatsoever. And since mathematics does not attempt to capture a representation of the world, a ``minimal interpretation'' always hides a particular ``metaphysical interpretation''. Unfortunately, there is in the literature such a ``minimal interpretation'' accepted ---either implicitly or explicitly--- by the vast majority of the community which, we will argue, is not only inconsistent but also closes the doors to a proper discussion and analysis of the metaphysical principles under consideration.

The most important metaphysical claim regarding the formalism of QM is found in the very first postulate of the theory which assumes the controversial fact that vectors describe `states of physical (quantum) systems'.\\

\noindent {\it {\bf POSTULATE I:} A vector in Hilbert space represents the physical state of a quantum system.}\\

\noindent This is indeed, the very origin of a strong metaphysical interpretation imposed on QM. As remarked by Michel Bitbol \cite[p. 72]{Bitbol10}: ``The tendency to reify state vectors manifests itself in the use of the very word `state'. The `grammar' (in Wittgenstein's sense) of the word `state' requires that this is the state of something; that it belongs to something; that it characterizes this something independently of anything else. Such grammar, and the conception associated to it, is sufficient to generate one of the major aspects of the measurement problem.'' Indeed, concepts allow us to think, they are not just words. 

Physical theories are constituted by a net of physical concepts, and it is these same concepts which allow us to configure physical experience, think about phenomena and even create new experiments. Classical mechanics, for example, was created not only through mathematical calculus but also through the notions of absolute space and time developed by Newton. Also the concept of field appears as necessary in order to understand Maxwell's electromagnetic theory. As Heisenberg makes the point \cite[p. 264]{Heis73}: ``The history of physics is not only a sequence of experimental discoveries and observations, followed by their mathematical description; it is also a history of concepts. For an understanding of the phenomena the first condition is the introduction of adequate concepts. Only with the help of correct concepts can we really know what has been observed.''

Nonetheless, in QM there is an accepted discourse used by the community, constituted by vaguely defined `common phrases', that has silently captured the theory within a specific metaphysical stance. For example, when discussing the interpretation, the orthodox view assumes that QM talks about ``quantum particles'', but the use of such concept limits the possibility itself of analysis, for the notion of ``particle'' is not empty of metaphysical commitments that we accept willingly or not when arguing under the presuppositions of such linguistic standpoint. Some, who know of the inconveniences of the use of such concept in order to refer to the formalism, argue that ``this is just a way of talking'', and immediately add: ``but we all know that QM does not talk about particles!'' We regard this contradiction in the use and meaning of language as very pernicious for a proper analysis and discussion of the semantics required in order to coherently interpret the theory. This is not ``just a way of talking" for the acceptance of a language implies constraints to the questions and problems that might be consistently posed. In particular, the notions of `state' and `system' have played a significant role in limiting the discussions regarding the interpretation of QM.

According to the orthodox interpretation, states of a quantum system are represented by normalized vectors of ${\cal H}$ and observables by self-adjoint operators $A$. The spectral theorem asserts that to any self-adjoint operator $A$ in an $n$-dimensional Hilbert space there exists an orthonormal basis $\{x_1,\ldots,x_n\}$ consisting of eigenvectors of $A$ such that:
$$A =\sum_{i=1}^n a_i|x_i\rangle\langle x_i|,\quad a_i\in\mathbb{R}.$$

\noindent The possible results of the measurement of a (sharp) magnitude are the eigenvalues $a_i$ of its associated operator $A$.  So observables may be decomposed to give an exhaustive and exclusive partition of the possible alternatives  for the results of measurements. The probability to obtain one of them in an experimental procedure is given by the Born rule. Given a vector $x\in H$, it is possible to write $x$ as a linear combination of the orthonormal basis $\{x_1,\ldots,x_n\}$, $$\sum_{i=1}^n \lambda_i|x_i\rangle,\quad \lambda_i=\langle x|x_i\rangle.$$

\noindent The coefficients $(\lambda_1,\ldots,\lambda_n)$ appearing in this expression are called the \emph{coordinates} of $x$ in the basis $\{x_1,\ldots,x_n\}$.

From a physical perspective, the notions of `state' and `system' used within QM imply the presupposition that we are discussing about an individual physical entity. This was indeed the first intuition of the early atomic theory which found its origin in the Democritean theory of atoms. As pointed out by Heisenberg:

\begin{quotation}
\noindent {\small``The strongest influence on the physics and chemistry of the past [19th] century undoubtedly came from the atomism of Democritos. This view allows an intuitive description of chemical processes on a small scale. Atoms can be compared with the mass points of Newtonian mechanics, and from this a satisfactory statistical theory of heat was developed. [...] the electron, the proton, and possibly the neutron could, it seemed, be considered as the genuinee atoms, the indivisible building blocks, of matter.'' \cite[p. 218]{Castellani98}}
\end{quotation}

\noindent At the beginning of the last century the founding fathers started their discussions about the meaning of such atoms (or quantum particles) but it soon became clear, that the notion of particle didn't coherently fit what the formalism was describing. Unfortunately, the first postulate is evidence of the paradoxical triumph ---even in QM--- of the metaphysics of objects and properties.

Of course, the notion of actual entity has been, since the origin of modern science, the main notion of physics. The notion of entity was conceived by Aristotle in terms of two realms of existence, potential and actual; the latter one described in terms of three logical and ontological principles: existence, non-contradiction and identity. But with the advancement of modern science only actuality remained as part of the physical realm. Newton was able to develop a mathematical formalism in order to deal with actual entities (see for discussion \cite{DFR06, RFD14b}). This development allowed to interpret a point in phase space as a physical object in (Newtonian) space-time. A physical system or entity was then conceived in terms of a set of definite valued properties. Even though one could observe such physical objects from different perspectives or reference frames the possibility to translate these views through the Galilean transformations secured the possibility of considering the measurement of subjects as external to a preexistent (objective) physical reality. Although it might be argued that this notion worked fairly well in order to develop physical theories, with the creation of QM many problems started to appear.

Indeed, the notion of actual entity has been a conceptual construction which allowed us to deal with experience for more than three centuries. Until QM, we could claim that all physics talked about some type of entity: particles, waves, rigid bodies, fields, etc. ---or even in more general terms, of an actual state of affairs. Every entity possesses a set of well defined objective ---meaning, subjective independent--- physical properties which can change through time. However, it is not obvious nor self evident that we should presuppose this specific `object-property' metaphysical scheme in order to interpret the quantum formalism. Unfortunately, the main discussions in the literature ---following implicitly the first postulate of QM--- presuppose implicitly the notion of object. This metaphysical choice has produced many interpretational problems. Obviously, such problems cannot escape their own presuppositions, and exactly because of this reason, QM has been confined to a discussion within the limits of this very specific metaphysical perspective. 

Since this discussion places the solution to the interpretation at the origin ---by assuming that we already know what QM talks about---, the analysis has only a negative perspective towards the theory. As a consequence the problems of QM are: non-locality, non-commutativity, non-distributivity, non-individuality, non-separability, etc.\footnote{We want to thank Bob Coecke for this linguistic characterization of the orthodox problems of QM presently discussed in the literature.} Even worse, forcing the metaphysics of objects with properties into the contextual quantum formalism has created the most weird mixture between subjective and objective aspects in a physical theory ever. As Janes makes the point:

\begin{quotation}
\noindent {\small``[O]ur present [quantum mechanical] formalism is not purely epistemological; it is a peculiar mixture describing in part realities of Nature, in part incomplete human information about Nature ---all scrambled up by Heisenberg and Bohr into an omelette that nobody has seen how to unscramble. Yet we think that the unscrambling is a prerequisite for any further advance in basic physical theory. For, if we cannot separate the subjective and objective aspects of the formalism, we cannot know what we are talking about; it is just that simple.''  \cite[p. 381]{Jaynes}}
\end{quotation}

\noindent We will argue that the orthodox interpretation of vectors as `physical systems' is not only hiding  a strong metaphysical commitment, but is, more importantly, simply inconsistent with the formalism of QM. A vector in Hilbert space escapes the structural constraints imposed by the metaphysical principles that determine the notion of physical entity. Thus the unscrambling of the subjective from the objective will have to do, first of all, with the abandonment of an improper metaphysical scheme dogmatically forced into the formalism of the theory. In the following section we will show that the invariance ---a main notion that allows physical theories to account for an objective description of reality--- that appears in QM is not consistent with the orthodox interpretation. The main point we will prove in this paper is that even though a vector is an invariant, it is not a mathematical invariant of the kind needed to provide an interpretation in terms of objects which possess physical properties.

\section{Classical Invariance and Quantum Contextuality}

In classical physics, every physical system may be described exclusively by means of its \emph{actual properties}, taking ``actuality'' as expressing the \emph{preexistent} mode of being of the properties themselves, independently of observation ---the ``pre'' referring to its existence previous to measurement. The evolution of the system may be described by the change of its actual properties. Mathematically, the state is represented by a point $(p; q)$ in the correspondent phase space $\Gamma$ and, given the initial conditions, the equation of motion tells us how this point moves in $\Gamma$. Physical magnitudes are represented by real functions over $\Gamma$. These functions can be all interpreted as possessing definite values at any time, independently of physical observation. Thus, as mentioned above, each magnitude can be interpreted as being actually preexistent to any possible measurement without conflicting with the mathematical formulation of the theory. In this scheme, speaking about potential or possible properties usually refers to functions of the points in $\Gamma$ to which the state of the system might arrive to in a future instant of time; these points, in turn are also completely determined by the equations of motion and the initial conditions. 

In QM, contrary to the classical scheme, physical magnitudes are represented by operators on ${\cal H}$ that, in general, do not commute. This mathematical fact has extremely problematic interpretational consequences for it is then difficult to affirm that these quantum magnitudes are \emph{simultaneously preexistent} (i.e., objective). In order to restrict the discourse to  sets of commuting magnitudes, different Complete Sets of Commuting Operators (CSCO) have to be (subjectively) chosen. In QM, contrary to the classical case, this (subjective) choice determines explicitly what is to be considered (objectively) real. This feature of the theory is known in the literature as {\it quantum contextuality}. Here is where the mixing of the objective and the subjective takes place. Indeed, the way to solve this uncomfortable situation within the orthodox perspective is to introduce a subjective choice ---in between the many contexts--- that reintroduces superficially the classical structure (see for discussion \cite{RFD14}). This {\it ad hoc} move which mixes the subjective with the objective has never been physically justified. Even the most accepted candidate to account for this interpretational maneuver, namely the principle of decoherence, has failed to provide a convincing physical explanation \cite{Adler03, Giulini96}. The idea that one needs to choose (subjectively) a context in order to determine which are the (objective) observables that have definite values violates explicitly counter factual reasoning  (see for example \cite{BeneDieks02}) which is maybe the most important feature of physical description itself ---a feature which allows us to go beyond the discourse about mere measurement outcomes. The violation of counter factual reasoning goes against the basic tenet of physical realism which claims that physics describes a world of which we humans are part, an objective world independent of consciousness or subjective human interventions. As a consequence, those who give up on counter factual reasoning also give up on the possibility of an objective physical description of reality. 

Starting from ``common sense'' classical realism instead of starting from the formalism of the theory ---which has already proven to be empirically adequate---, quantum contextuality has been considered in the literature as related to the problem of taking into account incompatible experimental arrangements. However, due to their invariance, vectors have seemed to escape this problematic situation. As a matter of fact, the first postulate of QM has remained a basic standpoint for the foundational problems and questions addressed within the literature. Indeed, most problems start their argumentation ---e.g. the PBR theorem, the problem of non-individuality, the problem of non-separability, the problem of non-locality--- assuming the validity of the first postulate. In this paper we will prove that such physical interpretation is inconsistent with the orthodox formalism of QM. This inconsistency has to do with a basic limitation of physical invariance within quantum theory. 

As remarked by Max Born \cite{Born53}: ``the idea of invariant is the clue to a rational concept of reality, not only in physics but in every aspect of the world.'' The notion of invariance allows us to determine what is to be considered {\it the same}. In physics, invariants are quantities having the same value for any reference frame. The transformations that allow us to consider the physical magnitudes from different frames of reference have the property of forming a group. In the case of classical mechanics we have the Galilei transformations which keep space and time apart, while in relativity theory we have the Lorentz transformations which introduce an intimate connection between space and time coordinates. Nomological properties are the main invariants which classify the physical system under study. In QM, nomological properties determine the classification of experience. Physicists classify elementary particles in terms of their {\it mass}, {\it spin} and {\it charge}, naming them according to their specific values as ``electrons'',  ``protons'', ``photons'', etc. However, the description provided by such nomological properties is completely static. Of course, what is really interesting for a physicist is not what remains always {\it the same} (independently of the observer) but rather what {\it changes:}  the dynamics. Restricting ourselves to physical magnitudes that remain always the same, independently of the reference frame, does not provide a dynamical picture of the world, instead such description only provides a static table of data. Obviously such description is completely uninteresting for physics, which attempts to describe not only how the world {\it is} but ---far more importantly--- how the world {\it changes}. 

That which matters the most for physical description is the {\it invariant variations} of physical magnitudes, that is, the dynamical magnitudes which vary but can be considered still the same (e.g., position, velocity, momentum, energy, etc.). {\it The difference within the identity.} Furthermore, in physics it is not only important to consider magnitudes that vary with respect to a definite reference frame ($S$), but also the consistent translation that allows us to consider that same variation with respect to a different frame of reference ($S'$). This relation (of the values between $S$ and $S'$) is also provided via the transformation laws. Such transformations include not only the dynamics of the observables but also the dynamics between the different observers. 
 
Even though the values of physical magnitudes might also vary from one reference frame to the other ---due to the dynamics between reference frames---, in both classical physics and relativity theory there is a {\it consistent translation} between the values of magnitudes of different frames secured by the transformation laws. The position of a rabbit running through the fields and observed by a distant passenger of a high speed train can be translated to the position of that same rabbit taken from the perspective of another passenger waiting in the platform of the station. The fact that the values of observables (position, momentum, etc.) can be consistently translated from one reference frame to the other allows us to assume that such physical observables also bear an objective real existence completely independent of the specific choice of the reference frame of the observes. The rabbit has a set of dynamical properties (position, a momentum, etc.) independently of his observers in the train and on the platform. The observables of the physical system are independent of the observers. We can thus claim that such properties are {\it dynamical variations} that pertain always to {\it the same} physical system. In more general terms, it is exactly this formal aspect which allows us to talk in terms of an {\it Actual State of Affairs (ASA)} that evolves in time; i.e., a dynamical description in terms of the variation of (objective) definite valued observables (or `dynamical properties') independent of the (subjective choice of the) perspective (or reference frame) from which they are being observed.\footnote{Even in relativity theory, due to the Lorentz transformations, one can still consider `events' as the building blocks of physical reality.} The same reasoning can be applied to coordinate transformations in the phase space $\Gamma$. If we consider a set of observables in a coordinate system, $S$, and perform a transformation of coordinates (e.g.,  a rotation) to a new system, $S'$, then the values of the observables will be also consistently translated from the system $S$ to the system $S'$. Such consistency, which is again secured by the transformation, is the {\it objectivity condition} which allows us to consider the observables as preexistent to the choice of the coordinate system.  

At a more profound level, apart from characterizing the evolution of systems, the dynamical properties are also responsible for allowing us to distinguish between different physical systems.\footnote{This can be related to the notion of imprimitivity. As remarked by Castellani \cite[p. 184]{Castellani98}: ``In the literature, the explicit use of the notion of imprimitivity system with regard to the definition of a `particle' is due especially to Piron \cite{Piron76}. The basic idea is to obtain a definition of the particle by employing physical quantities or {\it observables}, such as for example the position observable, through which the particle could be determined also as an individual object.''} As a matter of fact, without dynamical properties, physicists would not be even able to distinguish between physical systems that possess the same set of nomological properties. A physical description, without dynamics, would be a completely static table of data with no reference to particular physical systems or individuals. 

A physical system is necessarily described by its static (or nomological) properties and by its dynamical properties. As argued above, a necessary condition for claiming the reality of a physical system is that the valuation of such properties needs to be consistent with respect to different reference frames or coordinate transformations. Going now back to QM, since vectors are (by definition) invariants under rotations, it makes good sense at first sight to interpret a vector in Hilbert space as the state of an individual physical system. This is a very basic presupposition found within the orthodox postulates of QM. As we shall see, there is a subtlety involved in the formalism of QM which makes this interpretation untenable.

\section{A Corollary to Kochen-Specker Theorem}

In QM the frames under which a vector is represented mathematically are considered in terms of orthonormal bases. We say that a set $\{x_1,\ldots,x_n\}\subseteq {\cal H}$ in an $n$-dimensional Hilbert space is an \emph{orthonormal basis} if $\langle x_i|x_j\rangle=0$ for all $1\leq i<j\leq n$ and $\langle x_i|x_i\rangle=1$ for all $i=1,\ldots,n$. A physical quantity is represented by a self-adjoint operator on the Hilbert space ${\cal H}$. We say that $\mathcal{A}$ is a $\emph{context}$ if $\mathcal{A}$ is a commutative subalgebra generated by a set of self-adjoint bounded operators $\{A_1,\ldots,A_s\}$ in ${\cal H}$. Quantum contextuality, which was most explicitly recognized through the Kochen-Specker (KS) theorem \cite{KS}, asserts that a value ascribed to a physical quantity $A$ cannot be part of a global assignment of values but must, instead, depend on some specific context from which $A$ is to be considered. Let us see this with some more detail.

Physically, a global valuation allows us to define the preexistence of definite properties. Mathematically, a  \emph{valuation} over an algebra $\mathcal{A}$ of self-adjoint operators on a Hilbert space, is a real function satisfying,

\begin{enumerate}
\item \emph{Value-Rule (VR)}: For any $A\in\mathcal{A}$, the value $v(A)$ belongs to the spectrum of $A$, $v(A)\in\sigma(A)$.
\item \emph{Functional Composition Principle (FUNC)}: For any $A\in\mathcal{A}$ and any real-valued function $f$, $v(f(A))=f(v(A))$.
\end{enumerate}

\noindent We say that the valuation is a \emph{Global Valuation (GV)} if $\mathcal{A}$ is the set of all bounded, self-adjoint operators. In case $\mathcal{A}$ is a context, we say that the valuation is a \emph{Local Valuation (LV)}. We call the mathematical property which allows us to paste consistently together multiple contexts of {\it LVs} into a single {\it GV}, {\it Value Invariance (VI)}. First assume that a {\it GV} $v$ exists and consider a family of contexts $\{\mathcal{A}_i\}_I$. Define the {\it LV} $v_i:=v|_{\mathcal{A}_i}$ over each $\mathcal{A}_i$. Then it is easy to verify that the set $\{v_i\}_I$ satisfies the \emph{Compatibility Condition (CC)}, $$v_i|_{\mathcal{A}_i\cap \mathcal{A}_j}=v_j|_{\mathcal{A}_i\cap \mathcal{A}_j},\quad \forall i,j\in I.$$

\noindent The {\it CC} is a necessary condition that must satisfy a family of {\it LVs} in order to determine a {\it GV}. We say that the algebra of self-adjoint operators is \emph{VI} if for every family of contexts $\{\mathcal{A}_i\}_I$ and {\it LVs} $v_i:\mathcal{A}_i\rightarrow\mathbb{R}$ satisfying the \emph{CC}, there exists a {\it GV} $v$ such that $v|_{\mathcal{A}_i}=v_i$.

If we have {\it VI}, and hence, a {\it GV} exists, this would allow us to give values to all magnitudes at the same time maintaining a {\it CC} in the sense that whenever two magnitudes share one or more projectors, the values assigned to those projectors are the same from every context. The KS theorem, in algebraic terms, rules out the existence of {\it GVs} when the dimension of the Hilbert space is greater than $2$. The following theorem is an adaptation of \cite[Theorem 3.2]{DF} to the case of contexts:

\begin{theo}\label{CS2} {\bf (KS Theorem)} 
If ${\cal H}$ is a Hilbert space of $\dim({\cal H}) > 2$,
then a global valuation is not possible. \qed
\end{theo}

\noindent Let us analyze the case of a {\it LV}. Given a context $\mathcal{A}$ generated by a set of pairwise commuting self-adjoint operators $\{A_1,\ldots,A_s\}$ on ${\cal H}$, it is possible to find an orthonormal basis $\{x_1,\ldots,x_n\}$ of common eigenvectors to 
$\{A_1,\ldots,A_s\}$. Then, the context $\mathcal{A}$ is contained in 
the context generated by the operators $\{|x_i\rangle\langle x_i|\}_{i=1}^n$. We say that a context $\mathcal{A}$ is \emph{maximal} if
given a self-adjoint operator $B$ such that $BA=AB$ for all $A\in\mathcal{A}$, then $B\in \mathcal{A}$. Clearly, if $\mathcal{A}$
is maximal, then 
$\mathcal{A}$ is equal to 
the context generated by $\{|x_i\rangle\langle x_i|\}_{i=1}^n$.
The orthonormality implies that the operators $\{|x_i\rangle\langle x_i|\}_{i=1}^n$ are pairwise commuting and if a {\it LV} is defined
the \emph{Value-Rule} implies,
$$v(|x_i\rangle \langle x_i|)\in\{0,1\},\quad 1\leq i\leq n.$$
In particular, there are $2^n$ linear {\it LV} 
defined over a maximal context.

\

Let us prove that the group of rotations in QM does not preserve valuations in the same way as the transformation laws in classical mechanics and relativity theory. In classical mechanics and relativity theory one can predicate the truth or falsity of all the propositions with respect to the values of all (static and dynamic) physical magnitudes. The algebra of observables in both classical mechanics and relativity theory are commutative, hence the notion of {\it LV} and {\it GV} coincide ({\it LV} $\equiv$ {\it GV}). Even more, a change in the reference frame ---via a Galilean or a Lorentz transformation--- or a coordinate transformation in $\Gamma$ preserves the {\it CC}. This implies that we can assign a value to every observable simultaneously and in particular, we also have {\it VI}. This formal aspect of classical mechanics and relativity theory allows us to interpret physically both mathematical formalisms in terms of an {\it ASA}.

\begin{center}
\fbox{\parbox{1.6in}{
{\it VI} $\iff$ {\it GV} $\iff$ ASA}}
\end{center}

$VI$ can be also considered specifically with respect to nomological and dynamical properties. We can thus put forward the following two {\it necessary conditions} for considering observables as {\it objectively real properties of a physical system}: 

\begin{enumerate}
\item \emph{VI of Nomological Properties (VINP)}: The valuation of the set of nomological properties that constitute a physical system must be invariant under transformations of frames or coordinates.
\item \emph{VI of Dynamical Properties (VIDP)}: The valuation of the set of dynamical properties that constitute a physical system must be invariant under transformations of frames or coordinates.
\end{enumerate}

In QM, the algebra of observables is non-commutative. We will see that even though the invariance of nomological properties ($VINP$) is respected in the formalism, valuations of dynamical magnitudes are not preserved under rotations (failure of $VIDP$) and thus we do not have, in general, {\it VI} in QM. The physical consequence is that within the orthodox formalism of QM we cannot globally define a physical system nor an {\it ASA}. 

\

We say that the context $\mathcal{A}$ \emph{commutes} with
the context $\mathcal{B}$ if $AB=BA$ for all $A\in\mathcal{A}$
and $B\in\mathcal{B}$.
In particular, if $\mathcal{A}$ is maximal and
commutes with $\mathcal{B}$, then $\mathcal{B}\subseteq\mathcal{A}$
and any {\it LV} defined over $\mathcal{A}$ is defined 
over $\mathcal{B}$.
We have the following result,

\begin{theo}\label{teo-VI}
Let $v$ be a {\it LV} defined over a 
maximal context $\mathcal{A}$ and let $x \in {\cal H}$
be any vector. There exists a rotation of $x$ where $v$ is defined and
there exists a rotation where $v$ is not defined.
In particular, valuations are not preserved under rotations.
\end{theo}
\proof Assume that the dimension of ${\cal H}$ is greater that $2$.
Let $v$ be a local valuation defined over $\mathcal{A}$.
By Theorem \ref{CS2}, we know that there exists a vector $y\in {\cal H}$ such that $v$ is not defined over
$|y\rangle \langle y|$. By (FUNC) if $v$ is not defined over $|y\rangle \langle y|$, it is also not defined over
$|y'\rangle \langle y'|$,
$$y'=\frac{\|x\|}{\|y\|}y.$$
Given that $\|x\|=\|y'\|$, there exists a rotation sending $x$ to $y'$. Thus,
$v$ is not defined over a rotation of $x$. Similarly, we can rotate $x$ to a vector $z$
such that $v$ is defined over $|z\rangle \langle z|$. For example, 
take $z$ as a common eigenvector of the context $\mathcal{A}$
such that $\|z\|=\|x\|$.
\qed
\\

\noindent The previous theorem implies that under a rotation in 
${\cal H}$ the valuation is lost. Even though the vector $x \in {\cal H}$ is fixed, the coordinate system is fundamental in order to valuate $x$. We must choose another valuation or else the value of $x$ may not be defined.

We are now in conditions to state the following Corollary to KS Theorem:

\begin{coro}\label{CoroKS}{\bf}
If the dimension of the Hilbert space ${\cal H}$ is greater that $2$, then the $VIDP$ of a vector in ${\cal H}$ is precluded. 
\end{coro}
\proof
It follows from the failure of {\it VI}, Theorem \ref{teo-VI}.
\qed
\\

\noindent The main consequence of our corollary to KS theorem is that a vector in ${\cal H}$ cannot be interpreted as describing consistently dynamical properties. This precludes the interpretation of a vector in terms of an objectively real state of a physical system (or an $ASA$). 

\

Summing up, in QM there are no invariant variations or dynamical magnitudes which can be considered {\it the same} from different coordinate systems due to the failure of the $VIDP$ condition. As discussed above, only the dynamical properties are capable of providing a dynamical description of a physical system. What is most important in physics, change and variation, is not accounted for by nomological properties. Only nomological properties fulfill the objectivity condition with respect to vectors in ${\cal H}$. As we have discussed above this is a {\it necessary} but {\it not sufficient} condition to characterize a physical system. Also dynamical properties are abosultely necessary in order to define a physical system. Since it makes no sense to describe a system as possessing only nomological properties, the orthodox interpretation of a vector as representing the objective real state of a (quantum) system is inconsistent with the formalism of the theory. 

The corollary we derived from KS Theorem exposes an unfortunate common misuse of physical concepts when interpreting QM. In many regions of the literature such notions have produced many different pseudoproblems.

\section{KS Theorem and Quantum Individuality}

Since many foundational questions regarding the theory of quanta start their analysis and discussions taking as a standpoint the first postulate of the orthodox formulation our corollary to KS Theorem has consequences for many ongoing debates in the literature. For example, as Eugine Reich \cite[p. 157]{Reich12} explains ---making reference to the PBR theorem--- that Pusey, Barrett and Rudolph: ``say that the mathematics leaves no doubt that the wave function is not just a statistical tool, but rather, a real, objective state of a quantum system.''  This statement is false. As we have shown through corollary \ref{CoroKS}, the PBR theorem assumes a false hypothesis making its result untenable. The seemingly inoffensive assumption that a vector can represent the state of a physical system is simply inconsistent with the formalism of QM. 

Another important consequence of our theorem regards the idea that it makes perfect sense to consider, $\Psi_U$, ``the quantum wave function of the Universe'', or in other words, ``the Universal wave function''. This idea was first put forward by Hugh Everett in his Many Worlds (MW) interpretation of QM.  

\begin{quotation}
\noindent {\small``Since the universal validity of the state function description is asserted, one can regard the state functions themselves as the fundamental entities, and one can even consider the state function of the entire universe. In this sense this theory can be called the theory of the `universal wave function,' since all of physics is presumed to follow from this function alone.'' \cite[p. 9]{Everett73}}
\end{quotation}

\noindent As we have seen above, even if one would admit such idea, it becomes completely vague and unclear what kind of Universe we would be talking about. In particular,  corollary \ref{CoroKS} precludes the possibility that such Universe is represented or understood as an $ASA$. Even worse, since such Universe would be described consistently only by nomological properties, it would also be a completely static ---uninteresting--- Universe. But not only MW interpretation can be subject to strong criticisms from the perspective of our corollary \ref{CoroKS}, also Bohmian mechanics might be analyzed following these same lines of argumentation. As Michael Esfeld points out: 

\begin{quotation}
\noindent {\small``The Bohmian law for the temporal development of the distribution of matter in space is this one:

$$\frac{dQ}{dt} = v^{\Psi_t}(Q)$$

\noindent In this law, the quantum mechanical wave-function $\Psi$ has the job to determine the velocity of the particles at a time $t$, given their position at $t$. The wave-function can perform this job because it can with good reason be regarded as referring to a property, namely a dispositional property of the particles that determines their temporal development by fixing their velocity.

However, the wave-function that figures in [this] equation is the universal wave-function; consequently, $Q$ stands for the configuration of all the particles in the universe.'' \cite[pp 9-10]{Esfeld13}.}
\end{quotation}

\noindent However, a more careful analysis should be provided in this respect, for Bohmian mechanics changes the orthodox formalism of QM and might escape in this way the limits of the first postulate and the KS theorem altogether. 

Apart from theorems and interpretations, also some of the main problems of QM should be reconsidered taking into account what we have learnt. In this respect, a well known discussion in the literature which seems to be subject to the KS-criticism is the problem of non-locality, firstly suggested by Einstein through a thought experiment at the 1927 Solvay conference in Brussels. According to Einstein, if we accept that the quantum wave function provides a complete description of quantum particles then one can find that, after their interaction and when taken apart, QM predicts a ``spooky action at a distance'' between such quantum particles (see for discussion \cite{Berkovitz14}). As we have proven, the ``if'' assumed by Einstein is simply untenable in QM.\footnote{The fact that QM is not committed to this ``if'' was clearly acknowledged by Einstein himself, as clearly recognized in a letter to Born dated 5 April, 1948. See:  \cite[p. 168]{EinsteinBorn}.} The problem of non-locality rests on an assumption which is not consistent with the orthodox formalism. Notice that the problem of non-separability should be also reconsidered under this same KS type-criticism exposed in our corollary. 

Finally, we would like to remark that the problem of indistinguishable quantum particles, which is today one of the main foundational discussions in the literature, also takes for granted the first postulate. Within this debate, philosophers of physics have been arguing for and against about the distinguishability  and individuality of ``quantum particles''. Most part of this debate rests, as we have seen, on an untenable interpretation of the formalism. One cannot interpret a vector as being an actual individual entity. Due to failure of this interpretation the debate on identity and individuality should be critically reconsidered. We attempt to do so in a forthcoming paper.

\section{Final Remarks}

In this paper we have proven as a corollary to KS Theorem that the physical interpretation implied by the first postulate of QM is inconsistent with the orthodox formalism of the theory. In this respect, one might either still attempt to impose such metaphysical scheme to the formalism or simply accept that the notion of physical entity has been playing the role of an epistemological obstruction which avoids the possibility of understanding the theory through new physical concepts \cite{deRondeBontems11}. The assumption that QM talks about physical objects remains at this point of our knowledge ---more than one century after the creation of the theory--- simply a metaphysical presupposition that closes the door to a true development of the theory. In this respect, we might recall Einstein's considerations  with respect to the pernicious consequences of such dogmatic presuppositions regarding {\it a priori} concepts:

\begin{quotation}
\noindent{\small``Concepts that have proven useful in ordering things easily achieve such an authority over us that we forget their earthly origins and accept them as unalterable givens. Thus they come to be stamped as `necessities of thought,' `a priori givens,' etc. The path of scientific advance is often made impossible for a long time through such errors.'' \cite[p. 102]{Einstein16}}
\end{quotation}

According to the authors of this paper we should accept that the notions of `entity' and `actuality (as  a mode of existence)' have been creations, creations that impose a limit to physical representation, and that these limits seem to have been broken by QM. On the one hand, for those who accept the orthodox mathematics of QM, the use of concepts that do not match the formalism will only generate badly posed problems. On the other hand, the literature is full of new experimental and technological developments \cite{Nature13, NaturePhy12, Nature07} which should encourage us to develop new physical concepts that would allow us to represent and coherently think about such fascinating new phenomena. 

A mathematical formulation with no clear connection to physical concepts is only mathematics, it is not physics. This is why, from a realist stance, we are still in need to recover physical representation of QM and provide an answer to the question: what is the physical meaning of a vector in Hilbert space? We believe that the first step in order to find such a coherent interpretation should be to acknowledge which interpretations (of a vector in Hilbert space) are, due to the mathematical formalism, clearly precluded.

\section*{Acknowledgments}

The authors wish to thank Diederik Aerts, D\'ecio Krause, Ot\'avio Bueno and Newton da Costa for useful discussions on related subjects. They also want to thank Graciela Domenech, Hector Freytes and Jonas Arenhart for comments on earlier drafts of this paper. This work was partially supported by the following grants: VUB project GOA67 FWO project G.0405.08 and FWO-research community W0.030.06. CONICET RES. 3646/14 (2014-2015) and the Project PIO-CONICET-UNAJ (15520150100008CO) ``Quantum Superpositions in Quantum Information Processing''.


\begin{thebibliography}{10}

\bibitem{Adler03} Adler, S. L., 2003, ``Why Decoherence has not Solved the Measurement Problem: A Response to P. W. Anderson'', {\it Studies in History and Philosophy of Modern Physics}, {\bf 34}, 135-142.

\bibitem{BeneDieks02} Bene, G. and Dieks, D., 2002, ``A Perspectival
Version of the Modal Interpretation of Quantum Mechanics and the
Origin of Macroscopic Behavior'', {\it Foundations of Physics}, {\bf
32}, 645-671.

\bibitem{Berkovitz14} Berkovitz, J., 2014, ``Action at a Distance in Quantum Mechanics'', in {\it The Stanford Encyclopedia of Philosophy (Spring 2014 Edition)}, Edward N. Zalta (ed.), URL = http://plato.stanford.edu/archives/spr2014/entries/qm-action-distance/.

\bibitem{Nature13} Bernien, H., Hensen, B., Pfaff, W., Koolstra, G., Blok, M. S., Robledo, L., Taminiau, T. H., Markham, M., Twitchen, D. J., Childress, L. and Hanson, R., 2013, ``Heralded entanglement between solid-state qubits separated by three metres'',  {\it Nature}, {\bf 497}, 86-90.

\bibitem{Bitbol10} Bitbol, M., 2010, ``Reflective Metaphysics: Understanding Quantum Mechancis from a Kantian Standpoint", {\it Philosophica}, {\bf 83}, 53-83.

\bibitem{Born53} Born, M., 1953, ``Physical Reality", {\it Philosophical Quarterly}, {\bf 3}, 139-149.

\bibitem{EinsteinBorn} Born, M., 1971, {\it The Born-Einstein Letters}, Walker and Company, New York.

\bibitem{Castellani98} Castellani, E., 1998, {\it Interpreting Bodies. Classical and Quantum Objects in Modern Physics}, Princeton University Press, Princeton.

\bibitem{Conway} Conway, J., 1990, {\em A course in functional analysis}, volume~96 of {\em Graduate Texts in Mathematics}. Springer-Verlag, New York, second edition.

\bibitem{PS} Curd, M. and Cover, J. A., 1998, {\it Philosophy of Science. The central issues}, Norton and Company (Eds.), Cambridge University Press, Cambridge.

\bibitem{daCostadeRonde16} da Costa, N.C.A. and de Ronde, C., 2016, ``On the Applicability of Metaphysical Identity in Quantum Mechanics'', Preprint. 

\bibitem{deRondeBontems11} de Ronde, C. and Bontems, V., 2011, ``La notion d'entit\'{e} en tant qu'obstacle \'{e}pist\'{e}mologique: Bachelard, la m\'{e}canique quantique et la logique'', {\it Bulletin des Amis de Gaston Bachelard}, {\bf 13}, 12-38.

\bibitem{RFD14} de Ronde, C., Freytes, H. and Domenech, G., 2014, ``Interpreting the Modal Kochen-Specker Theorem: Possibility and Many Worlds in Quantum Mechanics'', {\it Studies in History and Philosophy of Modern Physics}, {\bf 45}, 11-18.

\bibitem{RFD14b}  de Ronde, C., Freytes, H. and Domenech, G., 2014, ``Quantum Mechanics and the Interpretation of the Orthomodular Square of Opposition'', in {\it New dimensions of the square of opposition}, Jean-Yves B\'eziau and Katarzyna Gan-Krzywoszynska (Eds.), Philosophia Verlag, Munich.

\bibitem{Dirac74} Dirac, P. A. M., 1974, {\it The Principles of Quantum Mechanics}, 4th Edition, Oxford University Press, London.

\bibitem{DF} Domenech, G. and Freytes, H., 2005, ``Contextual logic for quantum systems'', {\it Journal of Mathematical Physics}, {\bf 46}, 012102-1-012102-9.

\bibitem{DFR06} Domenech, G., Freytes, H. and de Ronde, C., 2006, ``Scopes and limits of modality in quantum mechanics", \textit{Annalen der Physik}, {\bf 15}, 853-860.

\bibitem{Einstein16} Einstein, A., 1916, ``Ernst Mach'' {\it Physikalische Zeitschrift}, {\bf 17}, 101-104.

\bibitem{Esfeld13} Esfeld, M., 2013, ``The reality of relations: the case from quantum physics'', in  {\it The metaphysics of relations}, A. Marmodoro and D. Yates (Eds.), Oxford University Press, Oxford, in press.

\bibitem{Everett73} Everett, H., 1973, {\it The Many Worlds Interpretation of Quantum Mechanics}, Princeton University Press, Princeton. 

\bibitem{Giulini96} Giulini, D., Joos, E., Kiefer, C., Kupsch, J., Stamatescu, I.-O., and Zeh, H. D., 1996, {\it Decoherence and the Appearance of a Classical World in Quantum Theory}, Berlin: Springer; second revised edition, 2003.

\bibitem{Heis58} Heisenberg, W., 1958, {\it Physics and Philosophy},
World perspectives, George Allen and Unwin Ltd., London.

\bibitem{Heis73} Heisenberg, W., 1973, ``Development of Concepts in
the History of Quantum Theory", in {\it The Physicist's Conception
of Nature}, 264-275, J. Mehra (Ed.), Reidel, Dordrecht.

\bibitem{Jaynes} Jaynes, E. T., 1990, {\it Complexity, Entropy, and the Physics
of Information}, W. H. Zurek (Eds.), Addison-Wesley.

\bibitem{KS} Kochen, S. and Specker, E., 1967, ``On the problem
of Hidden Variables in Quantum Mechanics", {\it Journal of
Mathematics and Mechanics}, {\bf 17}, 59-87. Reprinted in Hooker,
1975, 293-328.

\bibitem{NaturePhy12} Ma, X., Zotter, S., Kofler, J., Ursin, R., Jennewein, T., Brukner, C. and Zeilinger, A.,  2012, ``Experimental delayed-choice entanglement swapping'',  {\it Nature Physics}, {\bf 8}, 480-485.

\bibitem{Nature07} Ourjoumtsev, A., Jeong, H., Tualle-Brouri, R. and
Grangier, P., 2007, ``Generation of optical `Schr\"{o}dinger cats'
from photon number states'', {\it Nature}, {\bf 448}, 784-786.

\bibitem{Piron76} Piron, C., 1976, {\it Foundations of Quantum
Physics}, W.A. Benjamin Inc., Massachusetts.

\bibitem{PuseyBarrettRudolph12} Pusey, M. F., Barrett, J. and Rudolph, J., 2012, ``On the reality of the quantum state'',  {\it Nature Physics}, {\bf 8}, 475-478.

\bibitem{Reich12} Reich, E. S., 2012, ``A boost for quantum reality'',  {\it Nature}, {\bf 485}, 157-158.

\end{thebibliography}
\end{document}